\documentclass{svproc}
\usepackage{url}
\usepackage{graphicx}
\usepackage{verbatim}
\usepackage{tabularx}
\usepackage{hyperref}
\usepackage{booktabs}
\usepackage{cite}
\usepackage{float} 
\usepackage{amsmath}

\begin{document}
\mainmatter              
\title{An AI-Based Framework for Assessing Sustainability Conflicts in Medical Device Development}
\titlerunning{AI-Based Framework for Sustainability Conflicts}  
%
\author{Apala Chakrabarti}
%
\authorrunning{Apala Chakrabarti} 
%
%
\institute{Centre of Excellence in Design, \\
Department of Design and Manufacturing, \\
Indian Institute of Science, Bengaluru, India,\\
\email{apala@fsid-iisc.in}}

\maketitle              

\begin{abstract}
\par Designing sustainable medical devices requires balancing environmental, economic, and social demands, yet trade-offs across these pillars are difficult to identify using manual assessment alone. Current methods depend heavily on expert judgment, lack standardisation, and struggle to integrate diverse lifecycle data, which leads to overlooked conflicts and inconsistent evaluations. This paper introduces an AI-driven framework that automates conflict detection. Machine learning and natural language processing are used to extract trade-offs from design decisions, while Multi-Criteria Decision Analysis (MCDA) quantifies their magnitude through a composite sustainability score. The approach improves consistency, reduces subjective bias, and supports early design decisions. The results demonstrate how AI-assisted analysis provides scalable, data-driven support for sustainability evaluation in medical device development.

\end{abstract}
\section{Introduction}

\par The design of sustainable medical devices presents a unique challenge: design decisions often involve competing priorities across environmental impact, economic cost, and social responsibility. Balancing these dimensions is complicated by regulatory constraints, safety requirements, and market pressures specific to the medical device sector \cite{1}.

\par Recognizing sustainability conflicts, where progress in one area negatively impacts another, is crucial for effective decision-making. However, current approaches largely depend on expert judgment and assess pillars individually which leads to inconsistencies in identifying these conflicts \cite{2}

\par Artificial intelligence (AI), particularly large language models (LLMs), offers a promising avenue for automating sustainability analysis from unstructured design documentation. By embedding domain knowledge into prompt-based reasoning, LLMs can assist in parsing lifecycle narratives and detecting latent trade-offs.

\par This work explores the integration of LLMs into an existing sustainability conflict assessment framework, aiming to reduce reliance on expert labor and enable early-stage conflict detection. We propose two AI-assisted pipelines that leverage LLMs for life cycle stage classification and conflict identification, with outcomes evaluated using a structured MCDA-based scoring model \cite{10}.

\section{Background and Literature}

\par AI serves as a critical technological modality for advancing sustainability across diverse industrial and urban sectors. AI-driven systems leverage advanced computational paradigms, including machine learning (ML), predictive analytics, and optimization algorithms, to identify systemic inefficiencies, minimize waste generation, and optimize resource allocation, directly supporting the attainment of sustainable development goals (SDGs)\cite{3}.

\par Within complex supply chain operations, AI significantly enhances sustainability by optimizing logistical trajectories, thereby reducing fuel consumption, and facilitating granular, real-time emission monitoring. This capability concurrently ensures adherence to evolving regulatory compliance frameworks and preserves operational efficiency \cite{2,5}. Furthermore, the deployment of hybrid AI models, which integrate deep learning architectures with sophisticated optimization techniques, has demonstrated empirical evidence of substantial carbon footprint reductions, with reported decreases of up to 30\%, alongside notable improvements in overall resource efficiency within green logistics and supply chain paradigms \cite{3}.

\par Recent technical advancements underscore the efficacy of LLMs in sustainability assessment, particularly in urban planning contexts. LLMs enable the automation and standardization of project evaluation against established normative frameworks, such as ISO 37101. This facilitates consistent and rapid categorization of initiatives based on a multi-criteria sustainability assessment matrix \cite{4}. This methodological approach not only streamlines assessment procedures but also fosters a holistic understanding of project externalities, promoting integrated planning and decision-making. While AI offers extensive capabilities, the interpretation of its analytical outputs and the judicious consideration of ethical implications fundamentally necessitate human expertise \cite{4}.

\par Despite these advancements, prior work often demonstrates a fragmented approach to AI and natural language processing (NLP) techniques in sustainability-related domains. Specifically, current methodologies frequently address narrow tasks in isolation—such as classification, extraction, or summarization—without supporting comprehensive, design-phase reasoning. Moreover, the direct integration of LLMs into engineering contexts for holistic sustainability trade-off analysis and conflict detection remains largely unexplored. To address these limitations and enable more robust, lifecycle-level sustainability assessment, the following section introduces a novel methodology that leverages large language models for structured interpretation of engineering documentation, facilitates comprehensive sustainability assessment, and supports decision-making through conflict-aware analysis.

\section{Overall Methodology}

\par The product life cycle is divided into five stages: raw material acquisition, manufacturing, transportation, usage, and end-of-life. At each stage, cause-effect graphs are constructed to map design decisions to their impacts across the environmental, economic, and social pillars. A conflict is identified when a single cause results in both positive and negative effects across these pillars.

\par Identified conflicts are stored in a structured database. Each effect is assigned an impact score—High (0.75), Medium (0.5), or Low (0.25)—and weighted by the importance of its respective pillar. The sustainability score is calculated as the ratio of total negative impact to total impact:
\[
P = \sum_{i=1}^{n} \text{Impact}_{i}^{+} \times \text{Weight}_{\text{pillar}_i}
\quad ; \quad
N = \sum_{j=1}^{m} \text{Impact}_{j}^{-} \times \text{Weight}_{\text{pillar}_j}
\]
\[
T = P + N \quad ; \quad R = \frac{N}{T}
\]
where \(N\) is the total negative score, \(P\) is the total positive score, and \(T = P + N\) is the total weighted impact. The total impact score \(T\) also reflects the overall intensity of sustainability trade-offs and supports both product-level assessment and cross-product benchmarking.

\section{Proposed Frameworks}
\par This study presents two alternative NLP-based frameworks for sustainability conflict detection in medical device development. Both approaches employ a LLMS to analyze unstructured PLC documentation. The first framework operates through structured lifecycle segmentation and stage-wise conflict inference, while the second adopts an end-to-end model that directly extracts conflict statements from the text without explicit segmentation. 

\subsection{Framework A: Life-Cycle Staging and Scoring}
\par
\begin{figure}[H]
    \centering
    \includegraphics[width=0.7\linewidth]{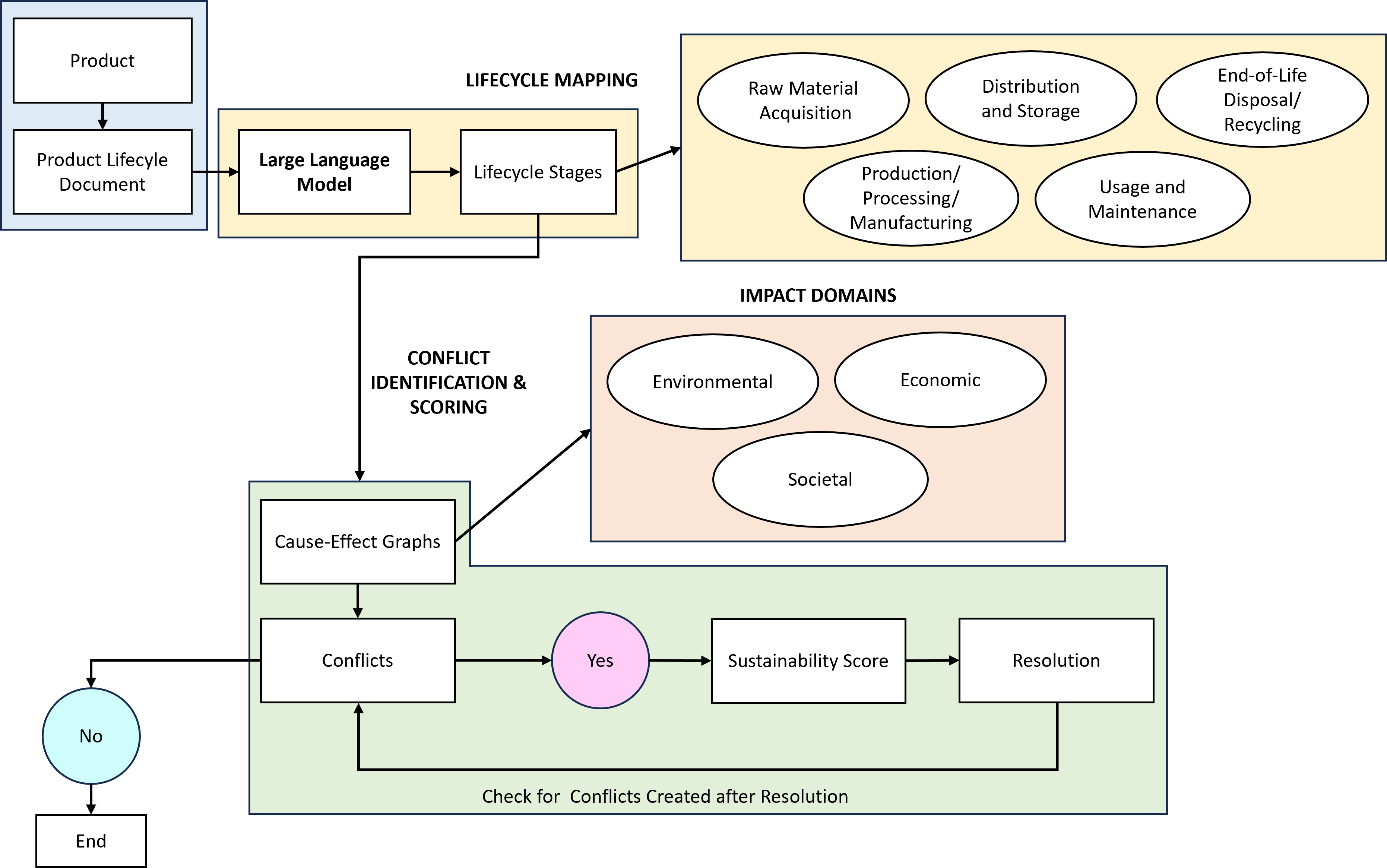}
    \caption{Process flowchart for sustainability assessment using Framework A}
    \label{fig:llm_1}
\end{figure}

 This framework incorporates an LLM to process the raw PLC document and divide it into five standardized lifecycle stages: raw material acquisition, production/processing/manufacturing, distribution and storage, usage and maintenance, and end-of-life disposal/recycling. For this implementation, the OpenAI API was used. Figure~\ref{fig:llm_1} illustrates the complete pipeline followed in this approach.

\par The LLM applies NLP techniques to classify sentences into appropriate lifecycle stages. Once segmented, the model supports CE mapping by associating design or operational decisions with sustainability impacts specific to each stage. These CE graphs are then used to infer sustainability conflicts by identifying mixed-polarity impacts across environmental, economic, and social pillars. Each conflict is then scored based on its severity and sustainability dimension.

\subsection{Framework B: Conflict Identification and Scoring}
\par Framework B extends the previous approach by introducing a second LLM to directly extract sustainability conflicts from the lifecycle-segmented text. As in Framework A, the input PLC document is first classified into five standard lifecycle stages. Subsequently, an additional LLM processes each stage to identify implicit and explicit sustainability trade-offs, producing structured conflict entries annotated with polarity and associated pillars.

\begin{figure}[H]
    \centering
    \includegraphics[width=0.7\linewidth]{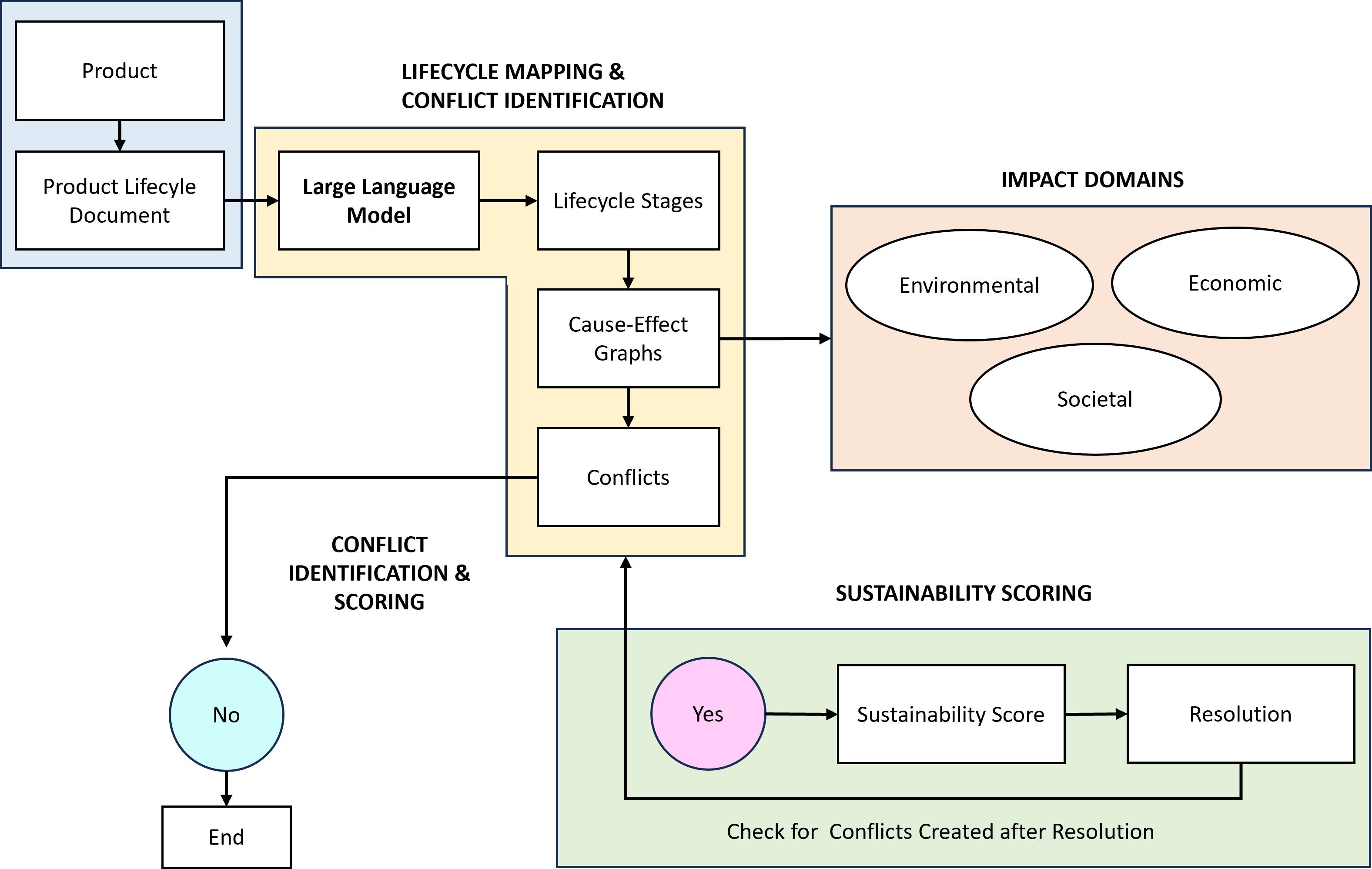}
    \caption{Process flowchart for sustainability assessment using Framework B}
    \label{fig:llm_2}
\end{figure}

\par This method preserves lifecycle context while automating conflict detection. As shown in Figure~\ref{fig:llm_2}, this two-stage pipeline enables scalable conflict extraction, particularly in documents with complex or diffuse sustainability information.

\section{Case Study: Mercury vs. Digital Thermometer}

\subsection{Manual Extraction}

\par Sustainability scores for both the mercury and digital thermometer was computed using the overall methodology. Table. \ref{tab:mt_manual_negative} and Table. \ref{tab:mt_manual_positive} reflect the 

\subsubsection{Mercury Thermometer (Manual)}

\begin{table}[H]
\vspace{-0.8em}
\caption{Positive sustainability effects: Mercury thermometer (manual)}
\label{tab:mt_manual_positive}
\begin{center}
\resizebox{0.6\linewidth}{!}{
\begin{tabular}{l c c}
\hline
\textbf{Effect ID} & \textbf{Sustainability Pillar} & \textbf{Impact Value} \\
\hline
MT-P1 & Environmental & 0.5 \\
MT-P2 & Environmental & 0.75 \\
MT-P3 & Economic & 0.25 \\
MT-P4 & Economic & 0.5 \\
MT-P5 & Economic & 0.75 \\
MT-P6 & Economic & 0.25 \\
MT-P7 & Economic & 0.5 \\
MT-P8 & Economic & 0.75 \\
MT-P9 & Social & 0.75 \\
MT-P10 & Social & 0.75 \\
\hline
\end{tabular}
}
\end{center}
\end{table}

\begin{table}[H]
\vspace{-0.8em}
\caption{Negative sustainability effects: Mercury thermometer (manual)}
\label{tab:mt_manual_negative}
\begin{center}
\resizebox{0.6\linewidth}{!}{
\begin{tabular}{l c c}
\hline
\textbf{Effect ID} & \textbf{Sustainability Pillar} & \textbf{Impact Value} \\
\hline
MT-N1 & Economic & 0.5 \\
MT-N2 & Economic & 0.75 \\
MT-N3 & Economic & 0.75 \\
MT-N4 & Economic & 0.75 \\
MT-N5 & Economic & 0.75 \\
MT-N6 & Social & 0.25 \\
MT-N7 & Social & 0.5 \\
MT-N8 & Social & 0.75 \\
\hline
\end{tabular}
}
\end{center}
\end{table}

MCDA scoring based on these extracted effects yields:  
\[
P = 0.5 + 0.75 + 0.25 + 0.5 + 0.75 + 0.25 + 0.5 + 0.75 + 0.75 + 0.75 = \textbf{5.75}
\]
\[
N = 0.5 + 0.75 + 0.75 + 0.75 + 0.75 + 0.25 + 0.5 + 0.75 = \textbf{5.0}
\]
\[
T = 10.75, \quad R = \frac{5.0}{10.75} \approx \textbf{0.465}
\]

\subsubsection{Digital Thermometer (Manual)}

\begin{table}[H]
\vspace{-0.8em}
\caption{Positive sustainability effects: Digital thermometer (manual)}
\label{tab:dt_manual_positive}
\begin{center}
\resizebox{0.6\linewidth}{!}{
\begin{tabular}{l c c}
\hline
\textbf{Effect ID} & \textbf{Sustainability Pillar} & \textbf{Impact Value} \\
\hline
DT-P1 & Environmental & 0.25 \\
DT-P2 & Economic & 0.25 \\
DT-P3 & Economic & 0.25 \\
DT-P4 & Economic & 0.5 \\
DT-P5 & Economic & 0.5 \\
DT-P6 & Economic & 0.75 \\
DT-P7 & Economic & 0.75 \\
DT-P8 & Economic & 0.75 \\
DT-P9 & Economic & 0.75 \\
DT-P10 & Social & 0.5 \\
DT-P11 & Social & 0.5 \\
DT-P12 & Social & 0.75 \\
DT-P13 & Social & 0.75 \\
\hline
\end{tabular}
}
\end{center}
\end{table}

\begin{table}[H]
\vspace{-0.8em}
\caption{Negative sustainability effects: Digital thermometer (manual)}
\label{tab:dt_manual_negative}
\begin{center}
\resizebox{0.6\linewidth}{!}{
\begin{tabular}{l c c}
\hline
\textbf{Effect ID} & \textbf{Sustainability Pillar} & \textbf{Impact Value} \\
\hline
DT-N1 & Environmental & 0.25 \\
DT-N2 & Environmental & 0.25 \\
DT-N3 & Environmental & 0.5 \\
DT-N4 & Environmental & 0.5 \\
DT-N5 & Environmental & 0.75 \\
DT-N6 & Social & 0.5 \\
DT-N7 & Social & 0.75 \\
DT-N8 & Social & 0.75 \\
\hline
\end{tabular}
}
\end{center}
\end{table}

MCDA scoring based on these extracted effects yields:  
\[
P = 0.25 + 0.25 + 0.25 + 0.5 + 0.5 + 0.75 + 0.75 + 0.75 + 0.75 + 0.5 + 0.5 + 0.75 + 0.75 = \textbf{7.0}
\]
\[
N = 0.25 + 0.25 + 0.5 + 0.5 + 0.75 + 0.5 + 0.75 + 0.75 = \textbf{4.25}
\]
\[
T = 11.25, \quad R = \frac{4.25}{11.25} \approx \textbf{0.378}
\]

\subsubsection*{Manual CE Graphs (Side-by-Side)}

\begin{figure}[H]
\centering
\begin{minipage}{0.48\textwidth}
  \centering
  \includegraphics[width=\linewidth]{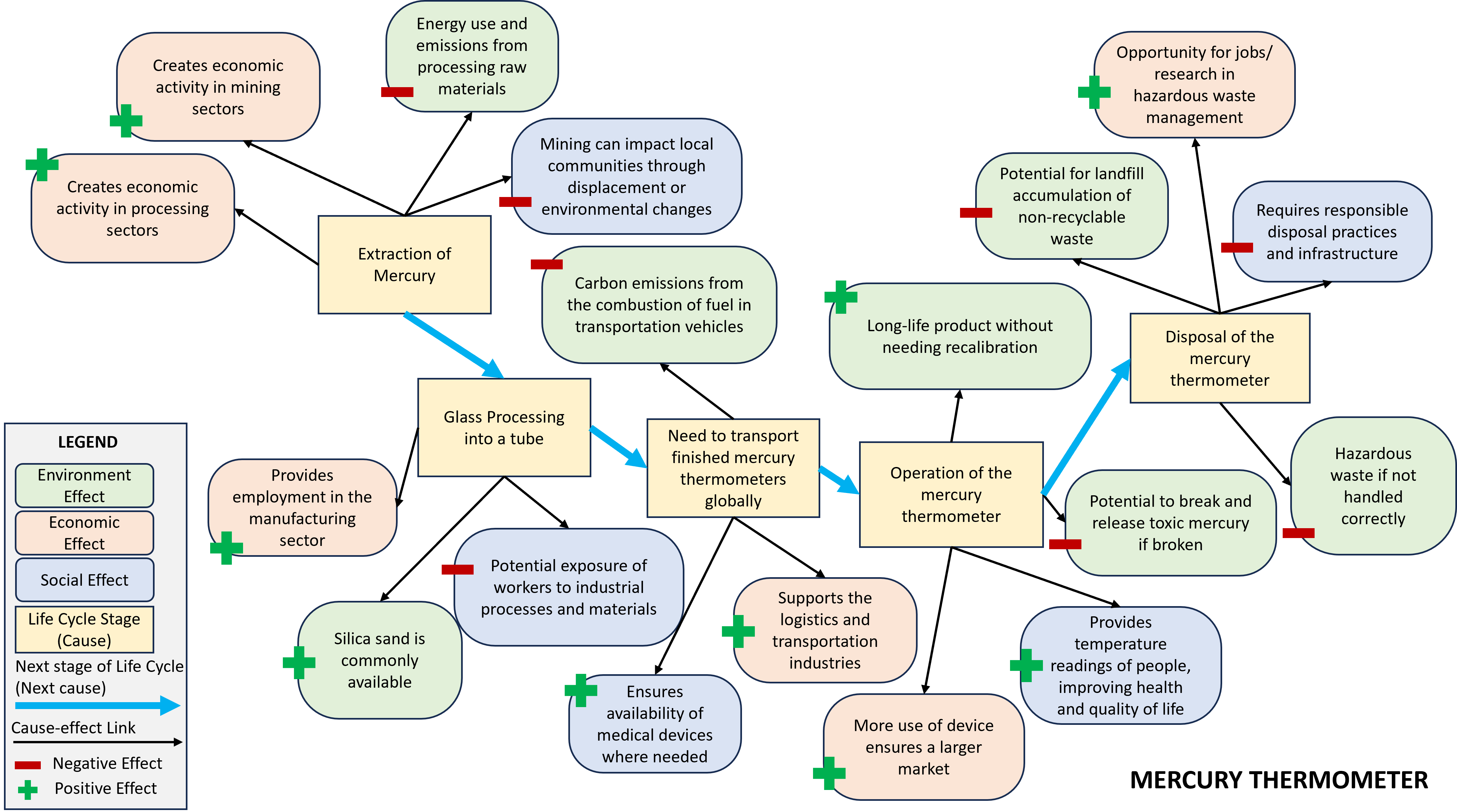}
  \caption*{Figure: Manual CE Graph for Mercury Thermometer}
\end{minipage}%
\hfill
\begin{minipage}{0.48\textwidth}
  \centering
  \includegraphics[width=\linewidth]{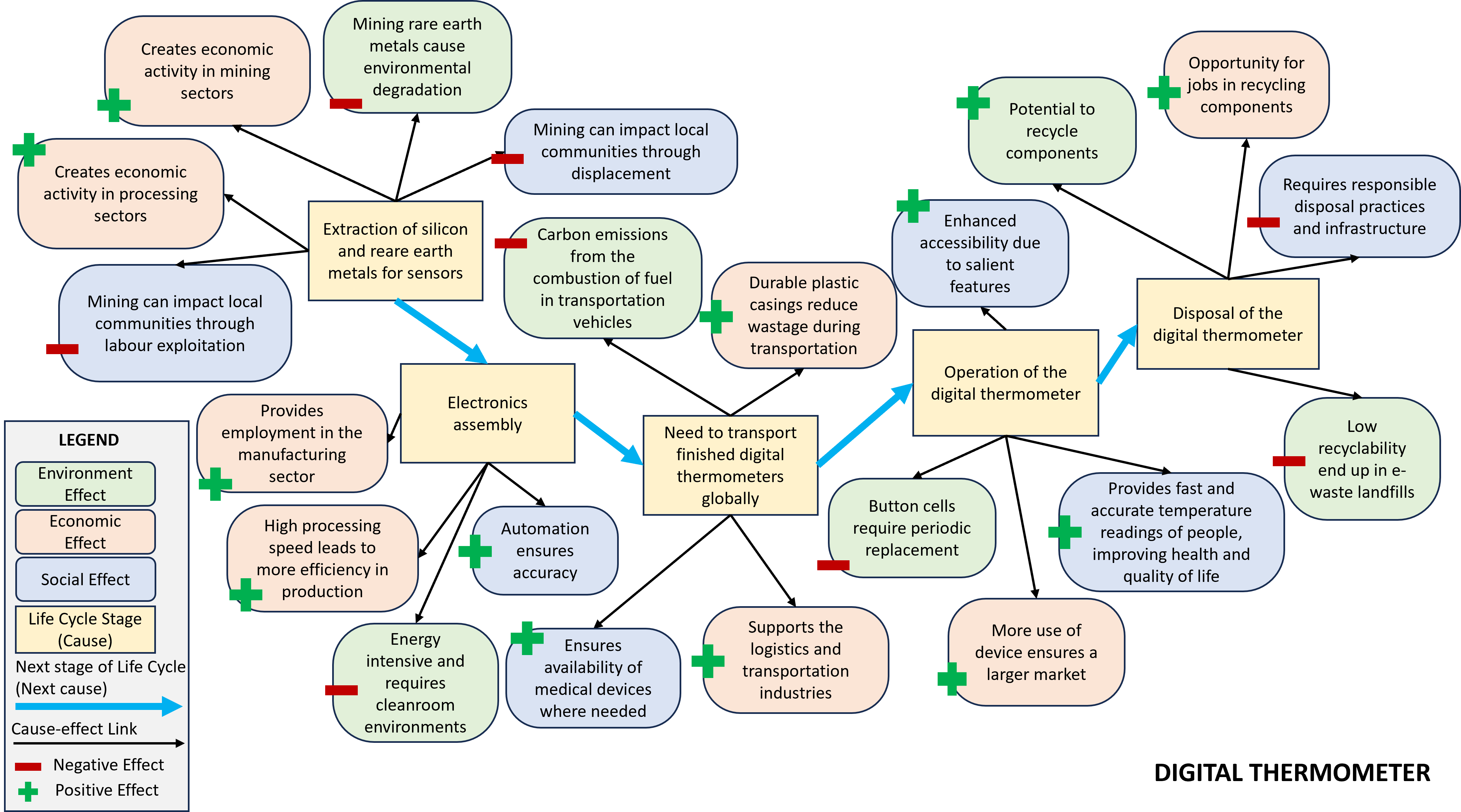}
  \caption*{Figure: Manual CE Graph for Digital Thermometer}
\end{minipage}
\end{figure}

\subsection{Framework A: NLP-Driven Conflict Extraction}

\par Framework A processes the PLC document by first segmenting it into five lifecycle stages using an LLM. Each sentence is parsed using SpaCy-based lemmatization and matched against curated keyword lists to infer three attributes: sustainability pillar (environmental, economic, social), impact severity (low, medium, high), and polarity (positive or negative). Unlike the manual method, \textbf{Framework A does not generate a cause-effect graph}; it directly extracts conflict-relevant sentences using heuristic filters and domain mappings.

\begin{table}[H]
\vspace{-0.8em}
\caption{Positive sustainability effects detected in the thermometer (Framework A)}
\label{tab:nlp_positive}
\begin{center}
\resizebox{0.6\linewidth}{!}{
\begin{tabular}{l c c}
\hline
\textbf{Effect ID} & \textbf{Sustainability Pillar} & \textbf{Impact Value} \\
\hline
CE-P1 & Environmental & 0.5 \\
CE-P2 & Environmental & 0.5 \\
CE-P3 & Environmental & 0.5 \\
CE-P4 & Social & 0.5 \\
CE-P5 & Environmental & 0.5 \\
\hline
\end{tabular}
}
\end{center}
\end{table}

\begin{table}[H]
\vspace{-0.8em}
\caption{Negative sustainability effects detected in the thermometer (Framework A)}
\label{tab:nlp_negative}
\begin{center}
\resizebox{0.6\linewidth}{!}{
\begin{tabular}{l c c}
\hline
\textbf{Effect ID} & \textbf{Sustainability Pillar} & \textbf{Impact Value} \\
\hline
CE-N1 & Economic & 0.5 \\
CE-N2 & Environmental & 0.75 \\
CE-N3 & Environmental & 0.5 \\
\hline
\end{tabular}
}
\end{center}
\end{table}

\par MCDA scoring based on these extracted effects yields:
\[
P = 1.375, \quad N = 0.75, \quad R = \frac{0.75}{2.125} \approx \textbf{0.353}
\]

\par The thermometer is classified as \textbf{Moderately Sustainable}. While Framework A provides structured lifecycle segmentation and automated output, it may overlook implicit or distributed trade-offs due to its reliance on rule-based keyword matching.

\section{Case Study 2: Sustainability of Eyewear Frames and Lenses}
\par To evaluate Framework B on a distinct medical device class, we applied it to eyeglasses, specifically comparing metal frames with glass lenses (MG) and plastic frames with plastic lenses (PG). Each variant underwent full 25-entry conflict extraction using the LLM pipeline, covering trade-offs across lifecycle stages and sustainability pillars. From the 25, we derived representative 15-entry sets, followed by 5-entry abstractions to retain high-level trends.

\par Positive and negative sustainability effects were tallied at each level and used to compute sustainability risk ratios using the LLM (Tables~\ref{tab:llm_scores}). A comparative CE graph was generated for the 5-entry set of each variant—both manually and via the LLM—to validate consistency and interpretability of the algorithmic outputs.

\begin{table}[H]
\vspace{-0.8em}
\caption{Sustainability Risk Ratios ($R = \frac{N}{P + N}$) for Metal and Plastic Eyeglasses (LLM-Generated)}
\label{tab:llm_scores}
\begin{center}
\resizebox{0.85\linewidth}{!}{
\begin{tabular}{lccc}
\hline
\textbf{Conflict Set Size} & \textbf{MG (Metal-Glass)} & \textbf{PG (Plastic-Plastic)} & $\Delta R$ (PG - MG) \\
\hline
5  & 0.528 & 0.591 & +0.063 \\
15 & 0.465 & 0.479 & +0.014 \\
25 & 0.436 & 0.475 & +0.039 \\
\hline
\end{tabular}
}
\end{center}
\end{table}

\section{Discussions, Conclusions and Future Work}

\par This paper proposed a structured framework to identify and quantify sustainability conflicts in medical device design. By integrating life cycle mapping, cause-effect analysis, and MCDA, the framework enables systematic evaluation across environmental, economic, and social pillars. The method allows differentiation of positive and negative sustainability effects, aggregation into a composite score, and structured interpretation of trade-offs. It supports holistic and transparent decision-making during early design stages.

Limitations include manual CE mapping, reliance on expert scoring, and limited scalability. Future work will focus on AI-assisted automation and exhaustive testing across device categories to enhance generalizability and adoption.

%
%

\end{document}